\documentclass{aa} 
\usepackage[utf8]{inputenc}
\usepackage{amsmath,amssymb}
\usepackage{graphicx} 
\usepackage{txfonts}
\usepackage{caption}
\usepackage{subcaption}
\usepackage{pifont}
\usepackage{float}
\usepackage{placeins}

\usepackage{hyperref}

\begin{document}

\title{Searching for X-Ray Counterparts of Degree Wide TeV Halos Around Middle-Aged Pulsars with SRG/eROSITA}

\titlerunning{Searching for X-Ray Counterparts of TeV Halos}

   \author{A. Khokhriakova
          \inst{1}
          \and
          W. Becker\inst{1,2}
          \and
          G. Ponti\inst{3,1}
          \and
          M. Sasaki\inst{4}
          \and
          B. Li\inst{5,6}
          \and
          R.-Y. Liu\inst{5,6}
          }

   \institute{
              Max-Planck Institut für extraterrestrische Physik, Giessenbachstraße, 85748 Garching, Germany
        \and
              Max-Planck Institut für Radioastronomie, Auf dem Hügel 69, 53121 Bonn, Germany
        \and
              INAF-Osservatorio Astronomico di Brera, Via E. Bianchi 46, I-23807 Merate (LC), Italy     
        \and
              Dr. Karl Remeis Observatory, Erlangen Centre for Astroparticle Physics, Friedrich-Alexander-Universität Erlangen-Nürnberg, Sternwartstraße 7, 96049 Bamberg, Germany
        \and
              School of Astronomy and Space Science, Nanjing University, Nanjing 210023, Jiangsu, China
        \and
              Key laboratory of Modern Astronomy and Astrophysics(Nanjing University), Ministry of Education, Nanjing 210023, People's Republic of China
              \\
        \email{alena@mpe.mpg.de}
             }

   \date{Received XXX; accepted YYY}

 
  \abstract
  {Extended gamma-ray TeV emission (TeV halos) around middle-aged pulsars has been detected. A proposed model to explain these TeV halos is that electrons from a degree-wide Pulsar Wind Nebula (PWN) get up-scattered by cosmic microwave background photons through inverse Compton processes. However, no X-ray degree-wide faint diffuse PWNe have been found around these middle-aged pulsars in previous X-ray observations.}
  {
    We have performed a search for degree wide PWNe around Geminga, PSR B0656+14, B0540+23, J0633+0632, and J0631+1036, using data from the first four consecutive Spectrum Roentgen Gamma/eROSITA all-sky surveys. In order to better understand the mechanisms underlying the formation of TeV halos, we investigated the magnetic field strength in the degree wide neighbourhood of those pulsars.
  }
  {
  To achieve our goals, we selected a list of suitable candidate pulsars in the eROSITA-DE part of the sky and applied data reduction techniques to process the eROSITA data. We then performed a spatial analysis of the regions around selected pulsars.
  } 
  {
  We did not detect degree-wide diffuse emission around Geminga, PSR B0656+14, B0540+23, J0633+0632, and J0631+1036, which can be attributed to being powered by the rotation-powered pulsars. Indeed, a close inspection of the data shows that the pulsars of interest are all embedded in diffuse emission from supernova remnants like the Monogem Ring or the Rosetta Nebula, while PSR B0540+23 is located $\sim 2.5$ degrees away from the bright Crab pulsar, which shines out the eROSITA point-spread function up to the position of PSR B0540+23 and thus reduced the sensitivity to search for degree wide bright diffuse X-ray emission strongly. 
  }
   {
   Despite the non-detection of any degree-wide PWN surrounding the analysed pulsars, we set flux upper limits to provide useful information on magnetic field strength and its spatial distribution around those pulsars, providing additional constraints to the proposed theory for the formation of TeV halos around pulsars.
   }


   \maketitle

\section{Introduction}

Recently, extended TeV emission (so-called TeV halos) has been detected around a few middle-aged pulsars. This includes the discovery of extended TeV emission around the nearby pulsars Geminga and PSR B0656+14 by the High Altitude Water Cherenkov observatory (HAWC) \cite{2017Sci...358..911A} as well as the source LHAASO J0621+3755 \cite{2021PhRvL.126x1103A} surrounding PSR J0622+3749 which was discovered by the Large High Altitude Air Shower Observatory (LHAASO). These findings strongly suggest that these phenomena may be common. A proposed model to explain this extended TeV emission and which relates it to the pulsars associated with it is that electrons from a degree wide Pulsar Wind Nebula (PWN) get up-scattered by cosmic microwave background (CMB) photons through inverse Compton processes \cite{2022MNRAS.513.2884L}.

These electrons can produce emission in the X-ray range via the synchrotron process in the interstellar magnetic field, which typically has a strength of 3 - 6 $\mu$G. 
Consequently, a diffuse X-ray emission with a comparable morphology to the TeV halos is expected \citep{2022MNRAS.513.2884L}.

Recently, X-ray emission has been studied in the region of the extended TeV source HESS J1809-193 \citep{2023arXiv230314946L}. 
The pulsar wind nebula of PSR J1809-1917, which is located within the extended gamma-ray emission, was a possible candidate for producing the TeV emission. 
The results show the presence of diffuse nonthermal X-ray emission extending beyond the PWN, which has been interpreted as an X-ray halo caused by escaping electron/positron pairs from the PWN. 
The analysis suggests that a relatively strong magnetic field of 21 $\mu$G is required to explain the spatial evolution of the X-ray spectrum \citep{2023arXiv230314946L}.

The High Energy Spectroscopic System (H.E.S.S.) is an array of Imaging Atmospheric Cherenkov Telescopes (IACT) 
\begin{figure*}
    \centering
    \includegraphics[width=\textwidth]{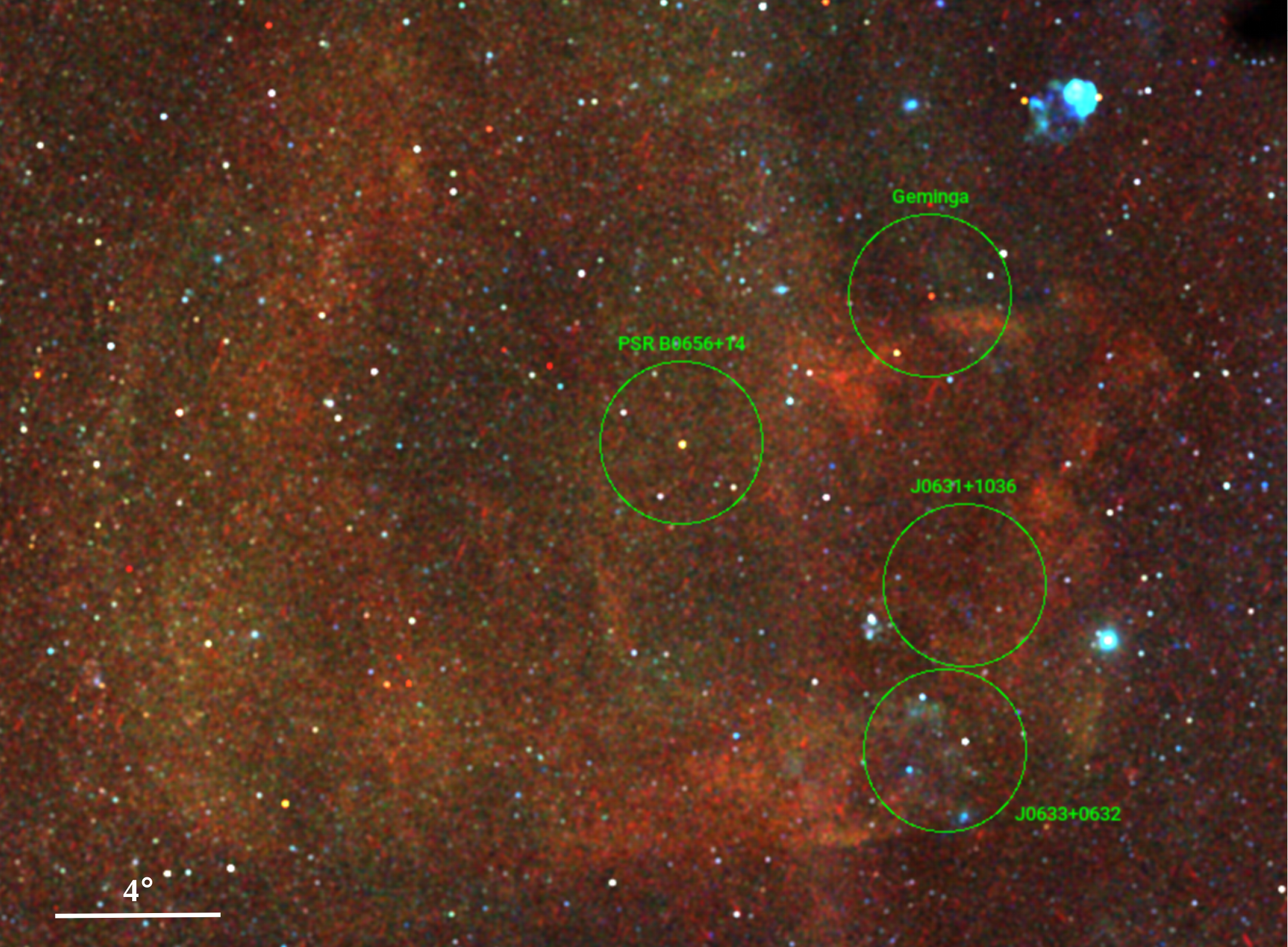}
    \caption{Image of the Monogem Ring complex as seen in the eROSITA 
    all-sky surveys eRASS:4 (cf. Knies et al., in prep.). { The supernova remnants IC443 and G189.6+03.3 (cf. Camilloni \& Becker, in prep.) are visible in the upper right corner.}
    Photons to produce the image were colour-coded 
    according to their energy (red for energies 0.2--0.4\,keV, green for 
    0.4--0.6\,keV, blue for 0.6--1.2\,keV). The green circles have a radius
    of 2 degrees centred on the position of the pulsars PSR B0656+14, 
    Geminga, PSR J0631+1036 and PSR J0633+0632.}
    \label{fig:Monogem}
\end{figure*}
\FloatBarrier
\noindent
designed to study cosmic gamma rays with energies ranging from $\sim$100 GeV to several tens of TeV \citep{2006ApJ...636..777A}. 
H.E.S.S. uses the imaging atmospheric Cherenkov technique to detect gamma rays via their interactions with the atmosphere, which produce a shower of charged particles that emit Cherenkov light.

The High Altitude Water Cherenkov Observatory (HAWC) 
uses the water Cherenkov detector array to detect gamma rays.
HAWC has been operating since 2015 and has a field of view of more than 2 sr and is sensitive to gamma rays in an energy range from hundreds of GeV to hundreds of TeV \citep{2020ApJ...905...76A}. 
The angular resolution of HAWC varies depending on the energy and zenith angle of the signal, ranging from 0.1$^{\circ}$ to 1.0$^{\circ}$. 

The Large High Altitude Air Shower Observatory (LHAASO) combines several different types of detectors:
1.3 km array of electromagnetic particle and muon detectors, a water Cherenkov detector array, 18 wide field-of-view air Cherenkov telescopes and an electron-neutron detector array.

Current X-ray telescopes, such as Chandra and XMM-Newton, have comparatively small fields of view and are therefore not very suitable for detecting such emission. { \citet{Liu2019_xray} analysed the data of Chandra and XMM-Newton around the Geminga pulsar and found no indication of the X-ray halo's emission. The result posed an upper limit of $0.6\,\mu$G for the interstellar magnetic field around the pulsar, or implies a mean magnetic field direction closely aligned with the line-of-sight of observer toward the pulsar \citet{Liu2019_ani}. However, the result only applies to the innermost 10' region from the pulsar due to the limited field of view of these two instruments.}  
Given the excellent sensitivity and unlimited field of view, eROSITA is an ideal instrument to search for the X-ray halo's emission or to deduce { more comprehensive constraints on the magnetic field in the halo region.} 

The eROSITA (extended ROentgen Survey with an Imaging Telescope Array) instrument \citep{2021A&A...647A...1P} is a space-based X-ray telescope designed to conduct a survey of the entire X-ray sky in unprecedented detail. 
Launched in 2019 as a part of the Russian-German { \it Spectrum-Roentgen-Gamma (SRG)} mission, eROSITA consists of seven identical X-ray mirror modules and by now completed four full all-sky surveys.

Excellent sensitivity in broad energy range ($\sim$ 0.2 - 10 keV) and large field of view (0.81 deg$^2$) make eROSITA a suitable 
\begin{figure*}
	\centering
	\begin{subfigure}[b]{0.45\textwidth}
		\centering
		\includegraphics[width=\textwidth]{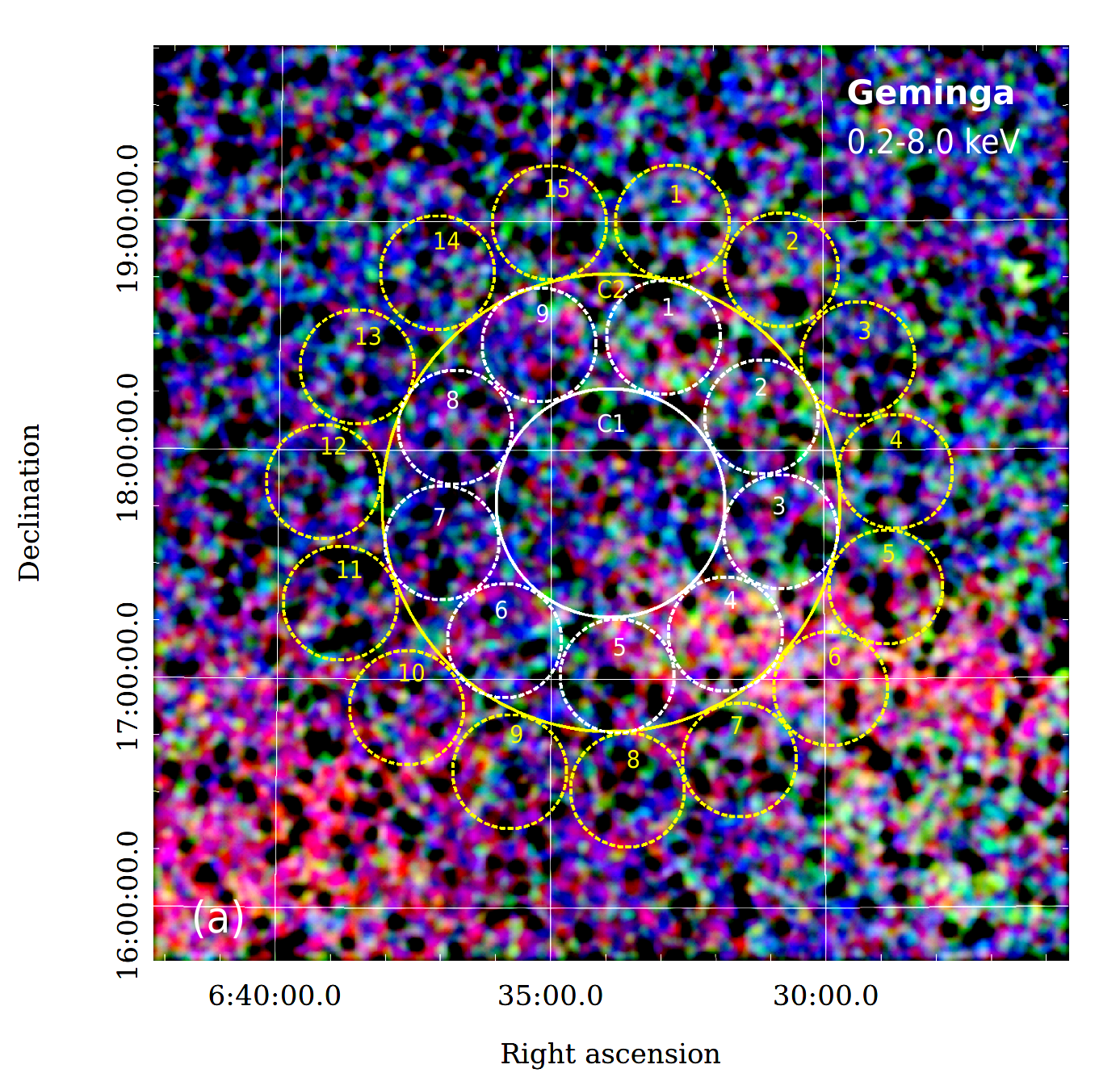}
		\label{fig:color-a}
	\end{subfigure}
        \quad
	\begin{subfigure}[b]{0.45\textwidth}
		\centering
		\includegraphics[width=\textwidth]{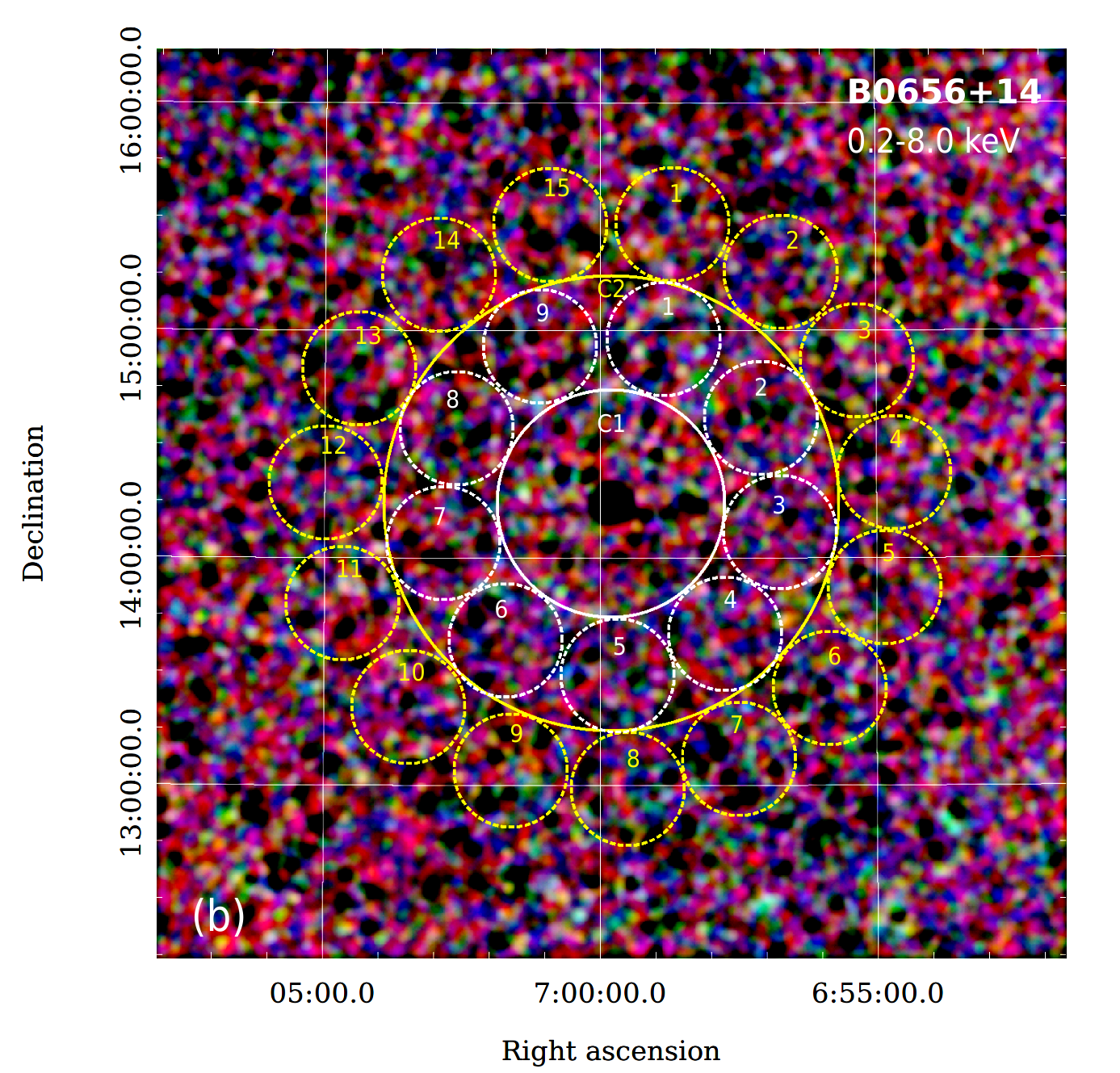}
		\label{fig:color-b}
	\end{subfigure}
        \\
        \begin{subfigure}[b]{0.45\textwidth}
		\centering
		\includegraphics[width=\textwidth]{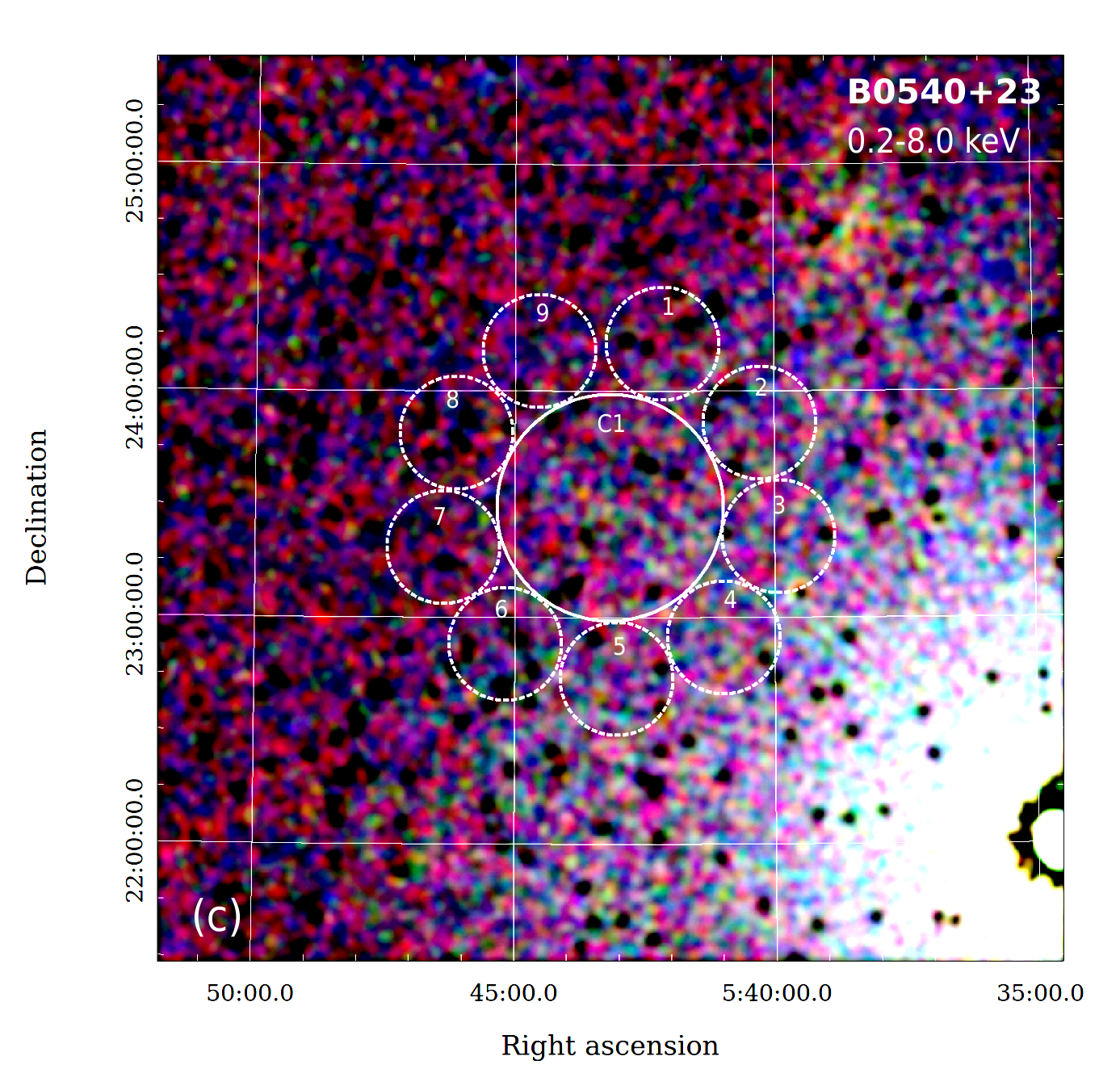}
	\end{subfigure}
    \quad
    \begin{subfigure}[b]{0.45\textwidth}
    \centering
    \includegraphics[width=\textwidth]{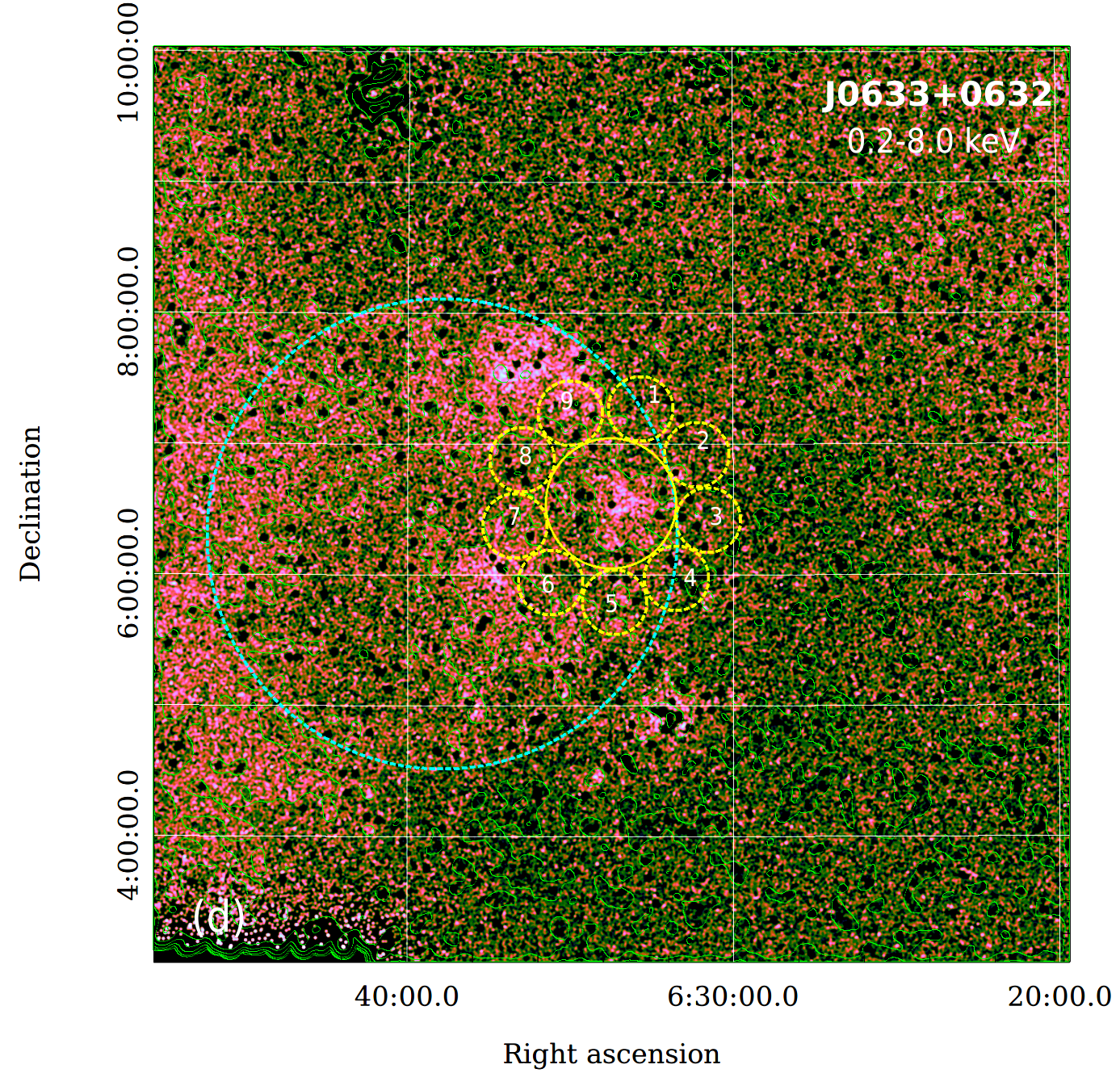}
    \label{fig:Monoceros}
\end{subfigure}
	\caption{The eROSITA X-ray images showing the regions surrounding the pulsars. (a)-(c): The RGB eROSITA X-ray images. Photons to produce the image were colour-coded according to their energy (red for energies 0.2–0.7 keV, green for 0.7–1.1 keV, blue for 1.1–8.0 keV). Point sources with a detection likelihood equal to or larger than 5$\sigma$ were removed. Solid circles have radii 0.5$^{\circ}$ (white) and 1$^{\circ}$ (yellow) and are used as source regions. Dashed circles with radii 0.25$^{\circ}$ are used as background regions. (d): The eROSITA X-ray image showing the region surrounding PSR J0633+0632. The blue circle indicates the Monoceros Nebula, which contributes to a higher background level in the regions 5-9.}
\label{fig:color}
\end{figure*}
\FloatBarrier
\noindent
instrument to measure the spatial distribution of the X-ray halos from sub-pc scale up to tens of pc scale.

eROSITA data are divided between German and Russian consortia. Data with Galactic longitude < 180$^{\circ}$ belong to the Russian consortium, while data with l > 180$^{\circ}$ (Western Galactic hemisphere) belong to the German eROSITA consortium (eROSITA-DE). 

In order to search for the X-ray halos, we selected a list of suitable candidate pulsars in the eROSITA-DE part of the sky. 
These include PSR J0633+1746 (Geminga), PSR B0656+14, PSR J0633+0632, PSR J0631+1036 and PSR B0540+23. 
However, { the first four pulsars} are located in the Monogem Ring, a nearby 20 degree wide supernova remnant, which hampers the search for faint diffuse emission around these pulsars (cf. Knies et al., in prep.). 
PSR B0540+23 is located near the bright Crab pulsar, which shines out the full eROSITA PSF, which partly overlaps the position of PSR B0540+23. 
Nevertheless, the detection of such emission or even upper limits of it may provide useful information on magnetic field strength and its spatial distribution around pulsars and gives important input into the proposed theory for the formation of TeV halos around middle-aged pulsars.

Our paper is organised as follows: In section 2, we describe the data reduction techniques used to process the eROSITA data. 
In section 3, we present the results of our spatial analysis of the regions around selected pulsars.
Section 4 summarises the results, including upper limits on the degree-wide diffuse emission around pulsars.
Finally, in section 5, we summarise our findings, highlighting the implications of the upper limits for the proposed model and including potential explanations for the lack of detection and suggestions for future research in this area. 

\section{Data Reduction}

We used data from the eROSITA telescope to search for X-ray halos around a sample of pulsars.
Our sample consisted of 5 pulsars, selected based on their high energy flux at Earth, age, magnetic field and the presence of TeV counterparts. 
Table~\ref{tab:candidates_prop} shows the
list of selected pulsars,
including those from \cite{2022MNRAS.513.2884L} and three additional pulsars.
It summarises the properties of the pulsars, including dispersion measure (DM), distance, age, surface magnetic field strength ($B_{\mathrm{surf}}$), spin-down power ($\dot E$), and energy flux at Earth ($\dot E/d^2$). 
The coordinates and properties are taken from the ATNF pulsar catalogue \citep{2005AJ....129.1993M}. 

\cite{2022MNRAS.513.2884L} argue that for pulsars to produce  X-ray halos, they should not be too young or too old.
Young pulsars are unlikely to exhibit halo-like emission since electrons accelerated by pulsar wind nebulae (PWNe) are injected into the interstellar medium (ISM) only a few tens of thousands of years after the pulsar's birth. On the other hand, the spin-down luminosity of pulsars diminishes with age, causing the diffuse X-ray emission surrounding old pulsars to be too faint to be detected. Therefore, X-ray halos are expected to be visible only around middle-aged pulsars, which have characteristic ages between 50 and 500 kyr. 

We expanded the sample from \cite{2022MNRAS.513.2884L} by including PSR J0631+1036 and PSR B0540+23. 
Although also the Crab pulsar satisfies the aforementioned criteria, it is too bright to apply a meaningful data analysis to it.
Moreover, it partly overlaps with the position of PSR B0540+23. The other four pulsars (Geminga, B0656+14, J0633+0632 and J0631+1036) are located in the Monogem Ring, a nearby 20-degree wide supernova remnant complex. Fig.~\ref{fig:Monogem} shows the Monogem Ring as seen by eROSITA during the four sky surveys. 
A detailed analysis of the Monogem Ring complex as seen by eROSITA { will be presented in an upcoming paper} by Knies et al. (in prep.). 
As can be seen from Fig.~\ref{fig:Monogem}, searching for faint diffuse emission around these pulsars is challenging because of the diffuse background contribution from the Monogem Ring itself.

\begin{table*}
\caption{Properties of X-ray halo candidate pulsars. Pulsar properties are taken from the ATNF pulsar catalogue \citep{2005AJ....129.1993M}.}
\label{tab:candidates_prop}
    \centering
    \begin{tabular}{l | cc ccc ccc}
        \hline\hline
        Pulsar name & R.A. & Decl. & DM & Dist. & Age & $\log_{10}B_{\mathrm{surf}}$ & $\log_{10}\dot E$ & $\log_{10}\dot E / d^2$\\
         & (deg) & (deg) & (cm$^{-3}$ pc) & (kpc) & (kyr) & (G) & (erg s$^{-1}$) & (erg s$^{-1}$ kpc$^{-2}$)\\
        \hline
        {  Geminga} & 98.47 & 17.77 & 2.89 & 0.19 & 342 & 12.212 & 34.505 & 35.947 \\ 
        {  B0656+14 } & 104.95 & 14.24 & 13.94 & 0.29 & 111 & 12.667 & 34.580 & 35.661 \\ 
        {  J0633+0632 } & 98.43 & 6.54 & n/a & 1.35 & 59.2 & 12.692 & 35.079 & 34.818 \\ 
        { J0631+1036 }  & 97.86 & 10.62 & 125.36 & 2.105 & 43.6 & 12.744 & 35.230 & 34.584 \\ 
        {  B0540+23 } & 85.79 & 23.48 & 77.70 & 1.57 & 253 & 12.294 & 34.613 & 34.223 \\
        \hline
    \end{tabular}
\end{table*}

To search for X-ray halos around pulsars, we analysed data from the eRASS:4 (stacked data from the first four consecutive all-sky surveys). 
Specifically, we used events in 0.2 - 8.0 keV energy range in regions around Geminga, B0656+14, J0633+0632, J0631+1036 ($\sim$400 s expositions ) and B0540+23 ($\sim$500 s exposition).

{ To combine data from all seven telescope modules (TMs) and merge the individual sky tiles, we followed the standard data reduction procedure using the \texttt{evtool} task provided in the latest internal release of eSASS, which is the eROSITA science analysis software \citep{2022A&A...661A...1B} (version \texttt{eSASSusers\_211214}). 
During the merging process, we employed the recommended flag and pattern filter keywords. To re-centre the images around the positions of the pulsars of interest, we utilised the \texttt{radec2xy} task. Subsequently, we created exposure-corrected images of the regions around these pulsars using the \texttt{expmap} task and binned them to a pixel size of $20^{\prime\prime}$ as per the established method. }

To avoid contamination, we masked out all highly significant point sources using the catalogue produced at MPE with eSASS.
We used detection likelihood threshold of \texttt{DET\_LIKE\_0} > 14.37 ($5 \sigma$ ) and the mask radius $120^{\prime\prime}$.
In the case of B0656+14, we masked the pulsar with a circle of $6^{\prime}$ radius because of its high brightness.

\section{Spatial Analysis}

In order to search for any potential extended X-ray emission around the pulsars, we first conducted a thorough visual inspection of the images obtained from eROSITA. Despite a careful examination, we did not detect any halo-like structures.

To obtain a quantitative measure of the significance of any potential extended emission, we extracted the source photons from circular regions with radii of 0.5$^{\circ}$ and 1$^{\circ}$ centred around the pulsars (see Fig.~\ref{fig:color}).
In order to determine the background levels, we employed 9 and 15 circular regions with radii of 0.25$^{\circ}$ around the smaller and larger source regions, respectively. 
We calculated the background level as the average of counts in these regions.
PSR J0633+0632 is located near the Monoceros Nebula (see Fig.~\ref{fig:color}d). Therefore, in this case, for background, we used only regions 5-9.

The uncertainty was estimated as 
\begin{equation}
    \sigma = \frac{\sqrt{N}}{t} = \frac{\sqrt{N}}{N} \times \text{CR},
\end{equation}
where $t$ is exposure, $N$ is the number of photons, CR is the count rate.
We also computed the resulting uncertainty in the background using error propagation:
\begin{equation}
    \sigma_{B} = \frac{\sqrt{\sum \sigma_i^2}}{n},
\end{equation}
where n is the number of background regions,
$\sigma_i = \frac{\sqrt{N_i}}{t_i}$ is the uncertainty of the $i$-th background region.

Our analysis indicates that there is no significant extended X-ray emission around the pulsars in the studied energy range.
Note that the background estimation is subject to potential systematic uncertainties due to the presence of contaminating sources.

Background variability can also be an important factor to consider. 
We investigated the background variability in the eROSITA data and found it to be rather constant. 
We examined light curves 
from background regions around all pulsars in our sample and observed no significant changes in the background signal over time within the errors. 

\section{Results}

Our analysis results are presented in Table~\ref{tab:results}, which displays the count rates and corresponding errors in 1-degree diameter circles and background regions for each selected pulsar. 
The first and second columns of the table provide the count rates and errors for the regions around the pulsar and the background, respectively.

Unfortunately, no diffuse emission was detected around the selected pulsars at a significance level of $3 \sigma$. 
To derive upper limits on the flux of the degree-wide emission, we calculated the maximum possible count rate as
\begin{equation}
    \text{CR}_{\text{max}} = (\text{CR}_{\text{s}} - \text{CR}_{\text{b}}) + 3 \times (\sigma_{\text{s}} + \sigma_{\text{b}})
\end{equation}
where $\text{CR}_{\text{s}}$ is the count rate in the central region,
$\text{CR}_{\text{b}}$ is the mean count rate in the background regions,
$\sigma_{\text{s}}$ and $\sigma_{\text{b}}$ are corresponding uncertainties.

Then, we converted the count rate to flux by assuming a power-law spectrum with a photon index $\Gamma = 2$ and simulating the observations using XSPEC.
The hydrogen column density ($N_{\mathrm{H}}$) was taken to be $2 \times 10^{20}$ cm$^{-2}$ for Geminga \citep{Mori_2014}, $1.7 \times 10^{20}$ cm$^{-2}$ for B0656+14 \citep{2022A&A...661A..41S}, $3 \times 10^{21}$ cm$^{-2}$ for J0633+0632 \citep{2015PASA...32...38D}, $2 \times 10^{21}$ cm$^{-2}$ for J0631+1036 \citep{Torii_2001}. For B0540+23, which was not detected in X-rays, we assumed $N_{\mathrm{H}} = 2 \times 10^{21}$ cm$^{-2}$ based on the known dispersion measure (DM) of 77.7 cm$^{-3}$ pc and the correlation between $N_{\mathrm{H}}$ and DM \citep{2013ApJ...768...64H}.
The resulting $3 \sigma$ flux upper limit for each pulsar is presented in the third column of Table~\ref{tab:results}.
{ We also examined the effect of varying parameters, such as the photon index and the size and shape of the background regions, on the derived upper limits. Our findings reveal that varying these parameters within a reasonable range ($\sim 15 \%$) yields consistent results, emphasising the stability of our upper limits.}

{
The magnetic field strength upper limits, presented in the fifth column of Table~\ref{tab:results}, have been determined by utilising the model introduced in \cite{2022MNRAS.513.2884L}. These upper limits were computed based on the observed X-ray upper limits associated with potential diffuse emissions surrounding the target pulsars, following the framework outlined in \cite{2022MNRAS.513.2884L}. 

Our analysis methodology involved fitting the TeV flux and surface brightness profiles while ensuring that the simulated X-ray flux remained below the eROSITA upper limits. For the TeV data, we relied on the TeV flux measurements from HAWC, as provided by \cite{2017Sci...358..911A}, for Geminga and B0656+14. For the remaining pulsars, we utilised LHAASO data \citep{2023arXiv230517030C}. 

To provide context for our findings, we compare them to the predicted interstellar magnetic field strengths around these pulsars, as calculated by the Jansson \& Farrar 2012 model (JF12) \citep{2012ApJ...761L..11J}. The derived magnetic field strength upper limit ($B_{\text{UL}}$) was found to be consistent with the expected interstellar magnetic field strength ($B_{\text{JF12}}$) around the first three pulsars in the table (Geminga, PSR B0656+14, and J0633+0632), while significantly lower limits were obtained for the last two pulsars (J0631+1036 and B0540+23).

It is worth noting that our findings provide important clues into the magnetic field geometry in the halo region under the anisotropic diffusion model of TeV halos, as described by \cite{2019PhRvL.123v1103L}. Additionally, we acknowledge the contribution of the study conducted by \cite{2019ApJ...875..149L}, in which magnetic field constraints around Geminga were obtained based on XMM and Chandra data, revealing an upper limit of $<0.6 , \mu\text{G}$ for the inner 10 arcminutes. The utilisation of eROSITA, with its broader field of view, enables a more comprehensive understanding of the magnetic field properties in these pulsar halos.

}

Fig.~\ref{fig:color} shows exposure-corrected images of the pulsars regions.
Point sources with detection likelihood corresponding to 5 $\sigma$ are removed. Solid circles have radii 0.5$^{\circ}$ (green) and 1$^{\circ}$ (yellow) and are used as source regions. Dashed circles with radii 0.25$^{\circ}$ are used as background regions.
The location of the PSR B0540+23 is very close to the bright Crab pulsar, which outshines the eROSITA PSF and makes analysing the background quite problematic. 

\begin{table*}
\caption[]{ X-ray count rates and flux upper limits (0.5-2.0 keV) along with magnetic field strength constraints}
    \centering
    \begin{tabular}{l |c c c c c}
        \hline \hline
        Pulsar name &  CR$_{\text{s}}$ & CR$_{\text{b}}$ & $F_{\text{UL}}$ & $B_{\text{UL}} $ & $B_{\text{JF12}}$\\
         & (cts s$^{-1}$ deg$^{-2}$) & (cts s$^{-1}$ deg$^{-2}$) & (erg s$^{-1}$ cm$^{-2}$ deg$^{-2}$) & ($\mu$G) & ($\mu$G)\\
        \hline
        Geminga & 13.51 $\pm$ 0.21 & 14.00 $\pm$ 0.14 & $4.69 \times 10^{-13}$ & $1.4$ & 1.6\\
        B0656+14 & 15.26 $\pm$ 0.23 & 14.13 $\pm$ 0.14 & $1.89 \times 10^{-12}$ & $4.0$ & 1.5\\
        J0633+0632 & 17.71 $\pm$ 0.23 & 17.27 $\pm$ 0.21 & $1.55 \times 10^{-12}$& $3.1$ & 1.5 \\
        J0631+1036 & 11.24 $\pm$ 0.19 & 11.37 $\pm$ 0.13 & $7.30 \times 10^{-13}$ & 2.6 & 8.0 \\
        B0540+23 & 17.85 $\pm$ 0.21 & 16.97 $\pm$ 0.14 & $1.73 \times 10^{-12}$ & 2.2 & 8.4 \\
        \hline
    \end{tabular}
    \label{tab:results}
\end{table*}

\section{Conclusion}

Our results show that no degree-wide diffuse halo-like emission was detected around the selected pulsars using eROSITA data. 
It can be attributed to several factors. 
In the case of PSR B0656+14, Geminga, PSR J0633+0632 and PSR J0631+1036, which are located inside the Monogem Ring, the diffuse emission from that SNR complex certainly reduces the sensitivity for detecting emission from low-surface brightness and degree wide plerions powered by the pulsars. Although the contribution from the Monogem Ring is mostly below 1 keV, we did not detect diffuse X-ray emission at higher energies (above 1 keV) as well. 

Similarly, PSR B0540+23 is located near the bright Crab pulsar, which shines out the full eROSITA PSF and partly overlaps the position of PSR B0540+23, making the detection of faint diffuse emission challenging.

Our upper limits on the degree-wide diffuse emission 
{ are } used to put constraints on the strength and spatial distribution of magnetic fields around pulsars. 
{ Our results show that the magnetic field strength in the TeV halos around Geminga, PSR J0631+1036 and PSR B0540+23 is lower than the anticipated average value of 3 $\mu$G \citep{2022MNRAS.513.2884L}. }

\begin{figure*}
	\centering
	\begin{subfigure}[b]{0.45\textwidth}
		\centering
		\includegraphics[width=\textwidth]{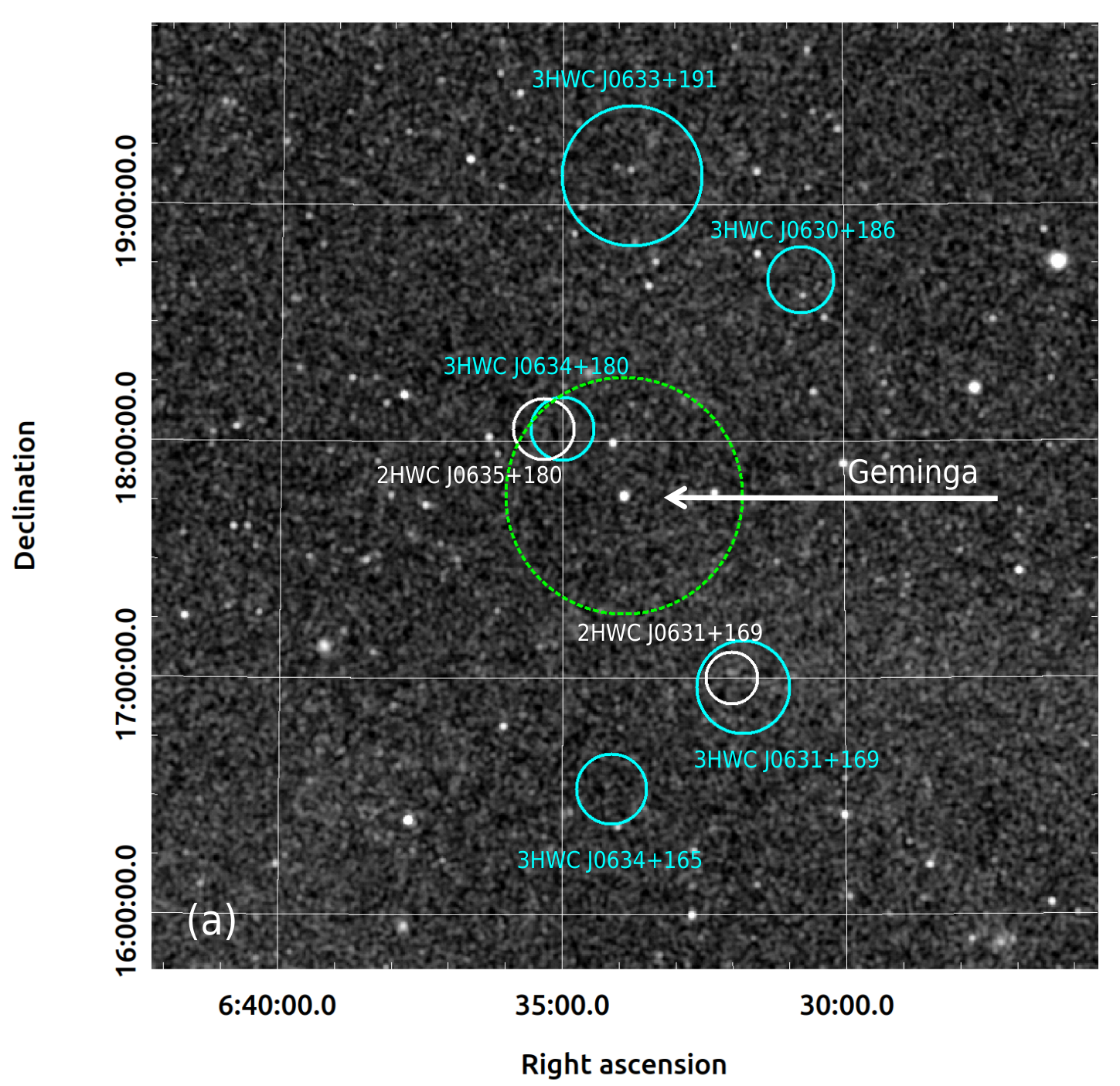}
		\label{fig:TeV-a}
	\end{subfigure}
        \quad
	\begin{subfigure}[b]{0.45\textwidth}
		\centering
		\includegraphics[width=\textwidth]{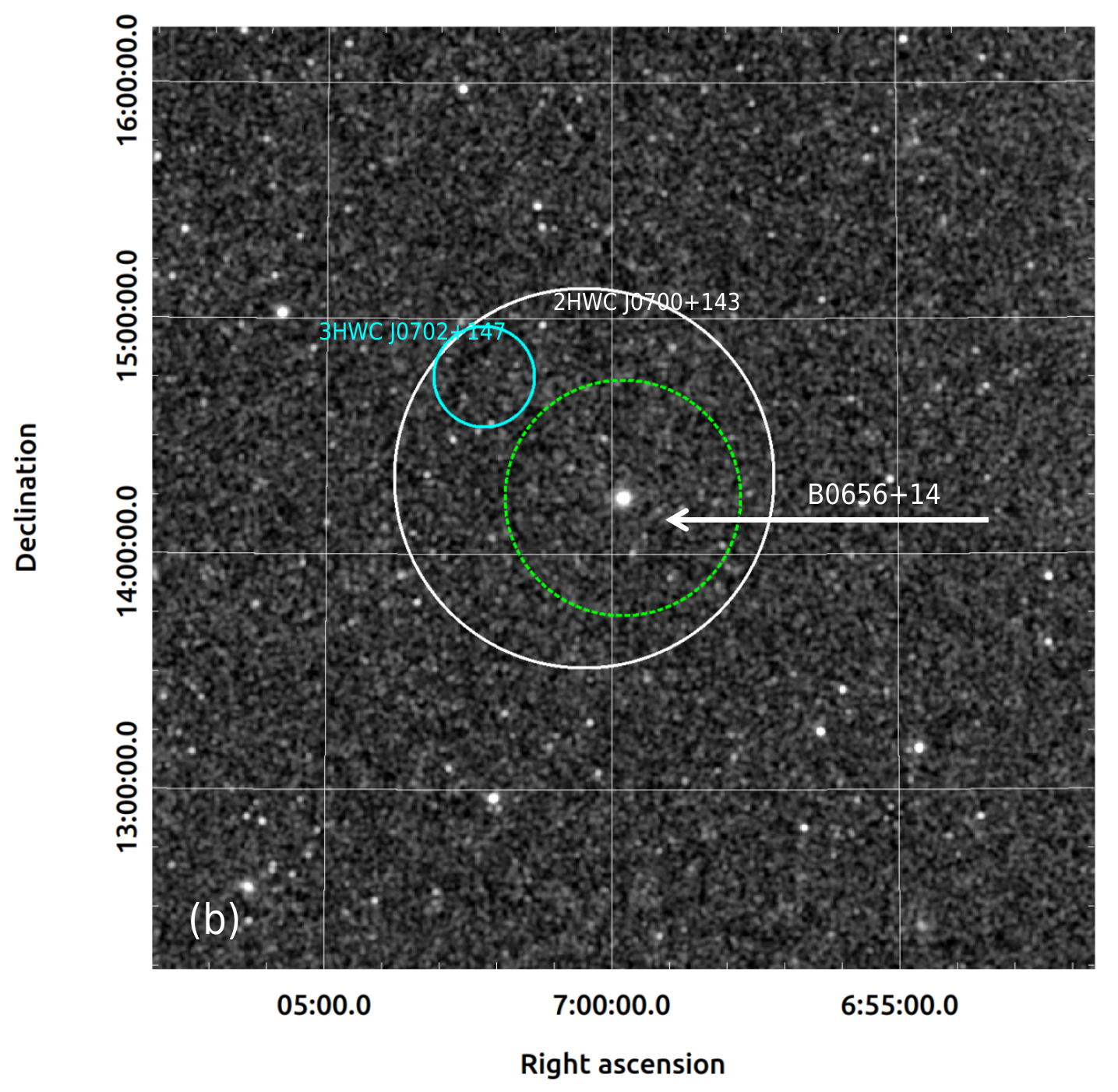}
		\label{fig:TeV-b}
	\end{subfigure}
        \\
        \begin{subfigure}[b]{0.45\textwidth}
		\centering
		\includegraphics[width=\textwidth]{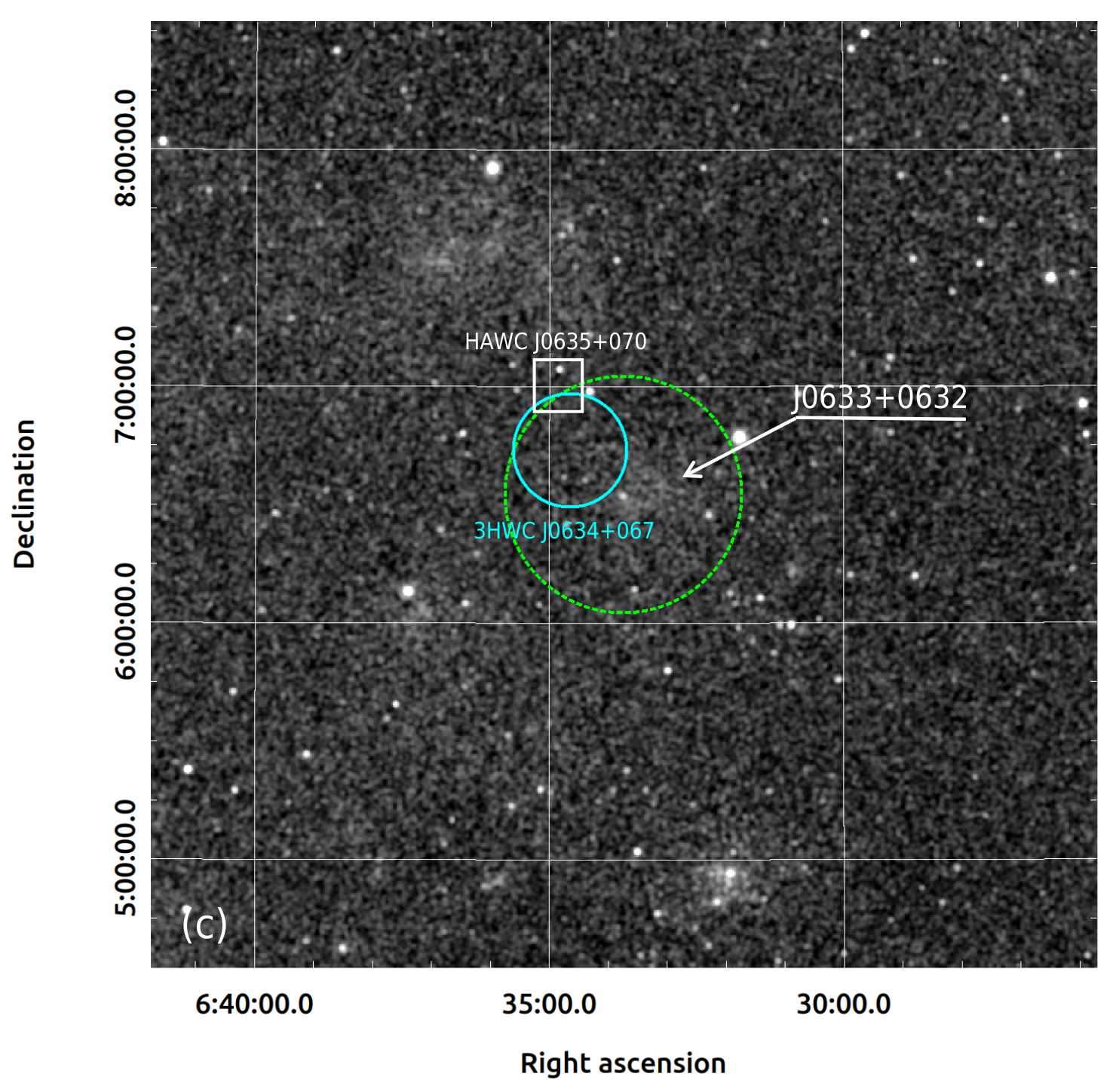}
		\label{fig:TeV-c}
	\end{subfigure}
        \quad
	\begin{subfigure}[b]{0.45\textwidth}
		\centering
		\includegraphics[width=\textwidth]{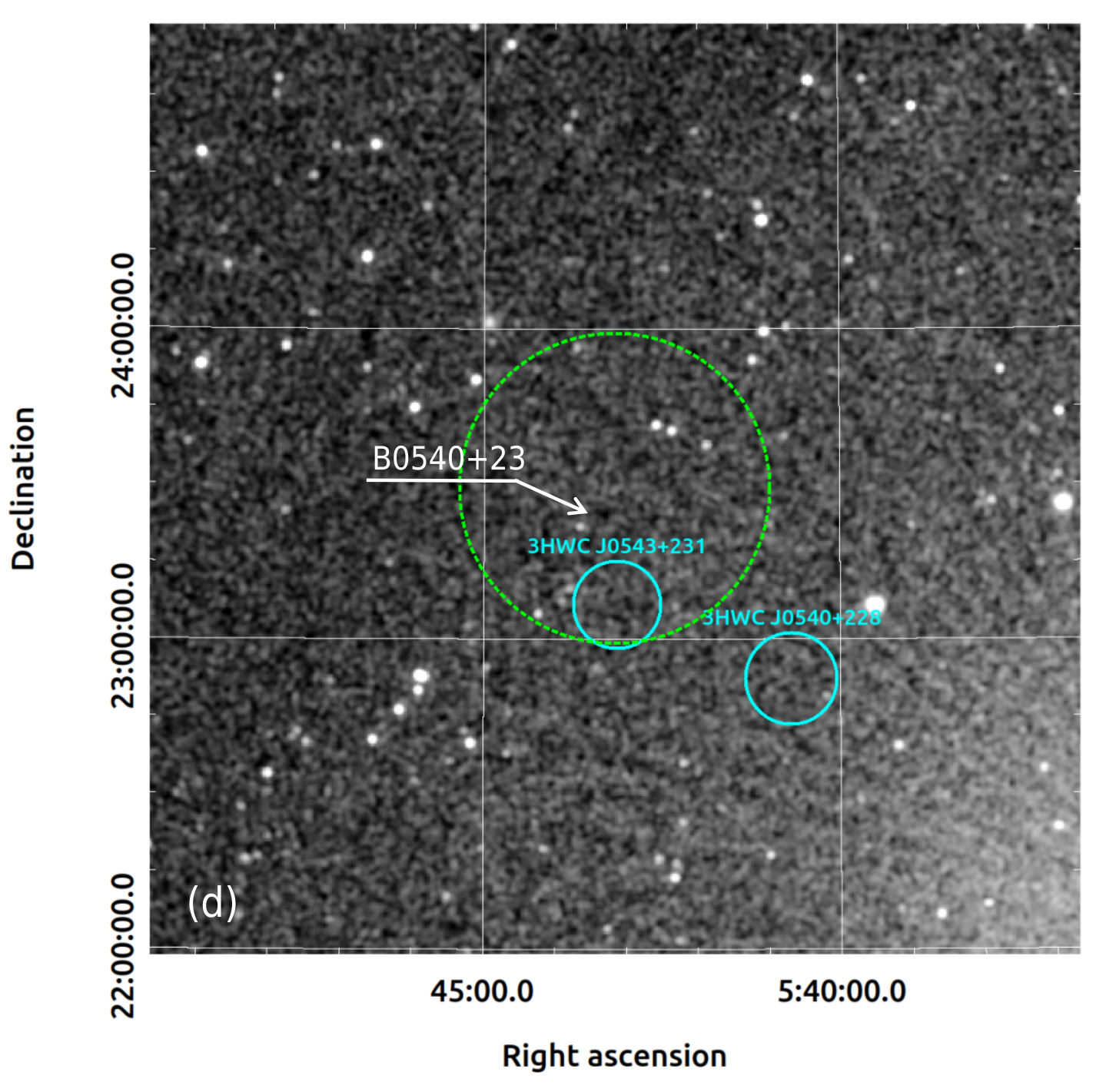}
		\label{fig:TeV-d}
	\end{subfigure}
	\caption{The eROSITA X-ray images showing the regions surrounding the pulsars. The white and blue circles and the square in Fig.c indicate the positions and uncertainties of the TeV sources from the 2HWC and 3HWC catalogues, respectively. The sizes of the circles correspond to the uncertainties of positions. The green dashed circles centred around the pulsars have a radius of 0.5 degrees. (a) Geminga. (b) B0656+14. (c) J0633+0632. (d) B0540+23.}
	\label{fig:TeV}
\end{figure*}

Alternatively, it is possible that for some pulsars, the TeV emission is produced by other sources in the vicinity of the pulsars.
Not all of the candidates are reliably associated with TeV sources. The positions of the possible TeV counterparts and corresponding uncertainties are shown in Fig.~\ref{fig:TeV}.
Only for B0656+14, the position of the pulsar lies within the error circle.

The $1\sigma$ positional error is computed as $\sqrt{r_c^2 + r_s^2}$,
where $r_c$ is $1\sigma$ statistical uncertainty taken from Table~1 in \cite{2020ApJ...905...76A},
$r_s = 0.1^{\circ}$ is the systematic pointing uncertainty.

Due to inhomogeneities in Galactic absorption, we do not provide analysis on a larger scale. Even on a scale of 1 degree, the background around the pulsars is highly inhomogeneous due to the presence of the nearby SNR Monogem Ring.

Overall, our study provides important insights into the magnetic field strength and spatial distribution around pulsars, contributing to a better understanding of the mechanisms underlying the formation of TeV halos.

\begin{acknowledgements}
 AK acknowledges support from the International Max-Planck Research School on Astrophysics at the Ludwig-Maximilians University (IMPRS).
 AK thanks Martin G. F. Mayer for the helpful discussions and suggestions.
 GP acknowledges funding from the European Research Council (ERC) under the European Union's Horizon 2020 research and innovation programme (grant agreement No 865637) and from "Bando per il Finanziamento della Ricerca Fondamentale 2022 dell'Istituto Nazionale di Astrofisica (INAF): Canale: GO Large program".
 This work is based on data from eROSITA, the soft X-ray instrument aboard \textit{SRG}, a joint Russian-German science mission supported by the Russian Space Agency (Roskosmos), in the interests of the Russian Academy of Sciences represented by its Space Research Institute (IKI), and the Deutsches Zentrum für Luft- und Raumfahrt (DLR). The \textit{SRG} spacecraft was built by Lavochkin Association (NPOL) and its subcontractors, and is operated by NPOL with support from the Max Planck Institute for Extraterrestrial Physics (MPE). The development and construction of the eROSITA X-ray instrument was led by MPE, with contributions from the Dr. Karl Remeis Observatory Bamberg \& ECAP (FAU Erlangen-Nuernberg), the University of Hamburg Observatory, the Leibniz Institute for Astrophysics Potsdam (AIP) and the Institute for Astronomy and Astrophysics of the University of Tübingen, with the support of DLR and the Max Planck Society. The Argelander Institute for Astronomy of the University of Bonn and the Ludwig Maximilians Universität Munich also participated in the science preparation for eROSITA. 
 The eROSITA data shown here were processed using the eSASS software system developed by the German eROSITA consortium. This work makes use of the Astropy Python package\footnote{\url{https://www.astropy.org/}} \citep{astropy:2013, astropy:2018, astropy:2022}. A particular mention goes to the in-development coordinated package of Astropy for region handling called Regions\footnote{\url{https://github.com/astropy/regions}}. We also acknowledge the use of Python packages Matplotlib \citep{Hunter:2007}
 and NumPy \citep{harris2020array}.
\end{acknowledgements}


\bibliographystyle{aa}
\bibliography{bib}

\onecolumn
\begin{appendix}

\section{eROSITA counts}

\begin{table*}[ht!]
\caption{eROSITA counts and effective areas for the regions around pulsars}
\label{tab:Geminga}
    \centering
\begin{tabular}{cccccc|cccccc}
\hline \hline
\multicolumn{6}{c|}{Geminga} & \multicolumn{6}{c}{B0656+14} \\
\hline
Region & Counts & Eff. Area & Region & Counts & Eff. Area & Region & Counts & Eff. Area & Region & Counts & Eff. Area \\
  &        & (deg$^2$) &   &        & (deg$^2$) &   &        & (deg$^2$) &   &        & (deg$^2$) \\
\hline
C1 & 9737 & 0.742 & C2  & 39980 & 2.970 & C1 & 10102 & 0.692 & C2  & 40907 & 2.871  \\
1 & 2620 & 0.191 & 1  & 2727  & 0.193 & 1 & 2692  & 0.183 & 1  & 2614  & 0.184 \\
2 & 2548 & 0.183 & 2  & 2473  & 0.186 & 2 & 2591  & 0.187 & 2  & 2536  & 0.184\\
3 & 2593 & 0.186 & 3  & 2352  & 0.182 & 3 & 2644  & 0.192 & 3  & 2644  & 0.179 \\
4 & 2974 & 0.192  & 4  & 2526  & 0.184 & 4 & 2672  & 0.182 & 4  & 2682  & 0.185 \\
5 & 2726 & 0.186 & 5  & 2533  & 0.175 & 5 & 2591  & 0.184 & 5  & 2660  & 0.185\\
6 & 2546 & 0.178 & 6  & 3174  & 0.198 & 6 & 2541  & 0.173 & 6  & 2543  & 0.175 \\
7 & 2167 & 0.182 & 7  & 2749  & 0.196 & 7 & 2459  & 0.180 & 7  & 2407  & 0.166 \\
8 & 2177 & 0.187 & 8  & 2458  & 0.189 & 8 & 2526  & 0.177 & 8  & 2484  & 0.178\\
9 & 2446 & 0.184 & 9  & 2388  & 0.189 & 9 & 2482  & 0.182 & 9  & 2410  & 0.170 \\
  &      &                   & 10 & 2500  & 0.185 &   &       &                   & 10 & 2756  & 0.1878\\
  &      &                   & 11 & 2191  & 0.173 &   &       &                   & 11 & 2625  & 0.176 \\
  &      &                   & 12 & 2251  & 0.187 &   &       &                   & 12 & 2532  & 0.182 \\
  &      &                   & 13 & 2222  & 0.183 &   &       &                   & 13 & 2477  & 0.178 \\
  &      &                   & 14 & 2427  & 0.189 &   &       &                   & 14 & 2585  & 0.192 \\
  &      &                   & 15 & 2278  & 0.182 &   &       &      & 15 & 2425  & 0.169\\
\hline
\hline
\multicolumn{6}{c|}{J0633+0632} & \multicolumn{6}{c}{J0631+1036} \\
\hline
Region & Counts & Eff. Area & Region & Counts & Eff. Area & Region & Counts & Eff. Area & Region & Counts & Eff. Area\\
  &        & (deg$^2$) &   &        & (deg$^2$) &   &        & (deg$^2$) &   &        & (deg$^2$) \\
\hline
C1 & 12076 & 0.754 & C2  & 44673 & 2.938 & C1 & 9191 & 0.720 & C2  & 37060 & 2.888  \\
1 & 2635  & 0.189  & 1  & 2550  & 0.189 & 1 & 1995 & 0.186   & 1  & 2149  & 0.186 \\
2 & 2618  & 0.180 & 2  & 2525  & 0.193 & 2 & 2426 & 0.188 & 2  & 2325  & 0.180 \\
3 & 2506  & 0.169 & 3  & 2518  & 0.183 & 3 & 2516 & 0.182 & 3  & 2639  & 0.191  \\
4 & 2415  & 0.186 & 4  & 2333  & 0.184 & 4 & 2389 & 0.157 & 4  & 2515  & 0.185 \\
5 & 2761  & 0.183 & 5  & 2438  & 0.191 & 5 & 2156 & 0.179 & 5  & 2642  & 0.187 \\
6 & 2902  & 0.189 & 6  & 2459  & 0.186 & 6 & 2327 & 0.186   & 6  & 2614  & 0.187 \\
7 & 2940  & 0.176 & 7  & 2469  & 0.187 & 7 & 2356 & 0.185 & 7  & 2496  & 0.177 \\
8 & 2662  & 0.182 & 8  & 2489  & 0.174 & 8 & 2333 & 0.192 & 8  & 2505  & 0.185 \\
9 & 3141  & 0.185 & 9  & 2908  & 0.183 & 9 & 2452 & 0.172 & 9  & 2639  & 0.193 \\
  &       &                   & 10 & 3092  & 0.1867 &  &      &                   & 10 & 2482  & 0.189 \\
  &       &                   & 11 & 3007  & 0.192 &   &      &                   & 11 & 2531  & 0.185 \\
  &       &                   & 12 & 2772  & 0.189 &   &      &                   & 12 & 2423  & 0.189 \\
  &       &                   & 13 & 3175  & 0.190 &   &      &                   & 13 & 2369  & 0.189 \\
  &       &                   & 14 & 3751  & 0.181 &   &      &                   & 14 & 2431  & 0.188 \\
  &       &                   & 15 & 2975  & 0.178 &   &      &                   & 15 & 2343  & 0.186 \\
\hline
\hline
\multicolumn{3}{c}{B0540+23} \\
\hline
Region & Counts & Eff. Area  \\
& & (deg$^2$) \\
\hline
C1 & 14136 & 0.751 \\
1 & 3633  & 0.183 \\
2 & 4165  & 0.190 \\
3 & 4225  & 0.193 \\
4 & 3854  & 0.187 \\
5 & 3105  & 0.171 \\
6 & 2887  & 0.193 \\
7 & 2611  & 0.176 \\
8 & 3024  & 0.190 \\
9 & 3086  & 0.184 \\
\hline
\end{tabular}
\tablefoot{Region C1 represents a 0.5 degree radius circle centred around a specific pulsar, while C2 corresponds to a 1 degree radius circle. Regions 1-9 and 1-15 refer to the background regions indicated in Fig.\ref{fig:color}. The effective area is calculated by subtracting the area of excluded point sources from the total region area. For PSR B0540+23, counts are provided only for the 0.5 degree circle due to contamination from the bright Crab pulsar (see Fig.\ref{fig:color}c).}
\end{table*}
    
\end{appendix}

\end{document}